\documentclass[review,number,sort&compress]{elsarticle}
\usepackage{lineno}
\usepackage{graphicx}
\usepackage{latexsym}
\usepackage{dcolumn}
\usepackage{bm}
\usepackage{makecell}
\setcellgapes{4pt}
\usepackage[english]{babel}
\usepackage[utf8x]{inputenc}
\usepackage[T1]{fontenc}

\newcommand{\ov}{$\mathrm{V_{OV}}$}
\newcommand*\DAlambert{\mathop{}\!\mathbin\Box}

\usepackage{amsmath}
\usepackage{graphicx}
\usepackage[colorinlistoftodos]{todonotes}
\begin{document}

\title{Cryogenic readout for multiple VUV4 Multi-Pixel Photon Counters in liquid xenon}

\author{F. Arneodo, M.L. Benabderrahmane, G. Bruno, V. Conicella, A. Di Giovanni, O. Fawwaz, M. Messina}
\address{New York University Abu Dhabi, Abu Dhabi, UAE}
\author{A. Candela}
\address{Laboratori Nazionali del Gran Sasso, Assergi AQ 67010, Italy}
\author{G. Franchi}
\address{Age Scientific SRL, Capezzano Pianore LU 55041, Italy}

\begin{abstract}
We present the performances and characterization of an array made of S13370-3050CN (VUV4 generation) Multi-Pixel Photon Counters manufactured by Hamamatsu and equipped with a low power consumption preamplifier operating at liquid xenon temperature ($\sim$ 175 K). The electronics is designed for the readout of a matrix of maximum dimension of  $\mathrm{8}\times\mathrm{8}$ individual photosensors and it is based on a single operational amplifier. The detector prototype presented in this paper utilizes  the Analog Devices AD8011  current feedback operational amplifier, but other models can be used depending on the application. A biasing correction circuit has been implemented for the gain equalization of photosensors operating at different voltages.
 The results show single photon detection capability making this device a promising choice for future generation of large scale dark matter detectors based on liquid xenon, such as DARWIN. 

\end{abstract}
\maketitle
\section{Introduction}

Liquefied noble gas targets are at the forefront of the search for dark matter \cite{ref:LUX,ref:XMASS,ref:DarkSide50}. In the upcoming generation of large scale detectors, a great emphasis will be given to compact photosensors suitable for cryogenic environment, with single photodetection response and allowing for large area coverage \cite{ref:Darwin}. A reduced radioactivity contribution to the total budget in order to minimize the experimental background is also crucial.
Direct detection of  vacuum ultraviolet (VUV) light  is  required by liquid xenon (LXe) based experiments ($\mathrm{\lambda_{scintillation}} \approx \mathrm{178~nm}$) \cite{ref:XENON}, while in liquid argon (LAr, $\mathrm{\lambda_{scintillation}} \approx \mathrm{125~nm}$) a wavelength shifter is usually needed \cite{ref:TPB}. A wavelength shifter is commonly used to shift the  $\mathrm{125~nm}$ scintillation light towards longer  wavelengths \cite{ref:TPB}. According to the most common WIMP models, the energy released in the interaction between dark and baryonic matter is supposed to be of the order of few tens of keV \cite{ref:Witten,ref:Wasserman}. 

To date, photomultiplier tubes (PMTs) are still the most widely used devices for scintillation light collection. 
In order to reach a more efficient coverage and to reduce the contribution of the  photosensors  to the total radioactivity budget of the detector, smaller devices  are being investigated. 
 With light yields in the liquid phase of the order of a few tens $\mathrm{photons/keV}$, to be able to achieve enough sensitivity, the detector must have a high geometrical coverage, single photon counting capability, adequate photon detection efficiency (PDE, larger than 20\% at the scintillation emission peak) and large gain (in the order of $\mathrm{10^6}$).  
A promising candidate is the Silicon Photomultiplier (SiPM) or Multi-pixel Photon Counter (MPPC) \cite{ref:MPPCworkingprinc}. 

\begin{figure}[h!]
\centering
\includegraphics[width=0.5\textwidth]{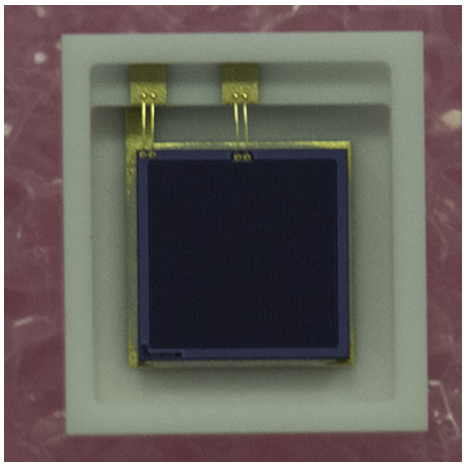}
\caption{\label{fig:S13370-3050CN} One of the $\mathrm{3 \times 3~mm^2}$ S13370-3050CN MPPCs used in the experiment.}
\end{figure}

The detector presented in this work is based on the use of the fourth generation of vacuum ultraviolet (VUV) multi-pixel photon counters (VUV4-MPPC) manufactured by Hamamatsu: the $\mathrm{3 \times 3~mm^2}$ S13370-3050CN, shown in figure \ref{fig:S13370-3050CN}.  

The most interesting features of the VUV4-MPPC are listed below: 

\begin{enumerate}
\item can be operated at low voltage ($\mathrm{<60~V}$) in LXe,
\item single photon counting capability,
\item PDE close to 25\% at $\mathrm{178~nm}$, 
\item gain larger than $\mathrm{2}\times\mathrm{10^6}$.
\end{enumerate}

To  offer the equivalent area of a standard  photomultiplier tube while keeping the same number of electronic channels, the grouping of several tens of  MPPCs is needed. This requirement poses a challenge in the design of the readout. A few typical readout examples are described in \cite{ref:MEGlayout}. More specifically, for a detector based on the use of a number $N$ of MPPCs, there are three configurations: 
\begin{itemize}
\item parallel of $N$ MPPCs,
\item series of $N$ MPPCs,
\item hybrid: parallels of two (or more) MPPCs connected in series.
\end{itemize}
The main characteristics of the possible configurations listed above are reported in Table \ref{tab:connections}.

\begin{table}[h]
\makegapedcells
\centering
\label{tab:connections}
\scalebox{0.9}{
\begin{tabular}{|c|c|c|c|c|c|}
\hline
$\mathbf{Configuration}$ & $\mathrm{Bias}$ & $\mathrm{Noise}$ & $\mathrm{Bandwidth}$ & $\mathrm{Gain~Uniformity}$ &$\mathrm{Coupling}$\\
\hline
$\mathbf{Parallel}$         &$\mathrm{1\times V_{bias}}$        & $\propto N\mathrm{\times C_s} $               & Low                    & Required& DC       \\
\hline
$\mathbf{Series}$         & $N\mathrm{\times V_{bias}}$        & $\propto \frac{\mathrm{C_s}}{N}$      & High                    & Automatic  &DC \\ 

\hline
$\mathbf{Hybrid}$         &$\mathrm{1\times V_{bias}}$        & $\propto 4\frac{\mathrm{C_s}}{N}$               & $\mathrm{\sim ~High}$       & Required& AC   \\ 
\hline
\end{tabular}}
\caption{Characteristics of $N$ MPPC based array for different readout configurations. {For the hybrid configuration we considered the series of groups of 2 MPPCs connected in parallel.}}
\end{table}
The noise contribution of each MPPC is due to its parasitic capacitance $\mathrm{C_s}$ and to the resistance 
$\mathrm{R_s}$ used for the connection to the operational amplifier inverting input. 

In the ideal case, for which the parasitic capacitance is extremely low ($\mathrm{C_s}\simeq 0)$, the contribution to the overall noise is given by the input voltage noise of the operational amplifier. In the real case, the contribution to the overall noise is given by the number of connected MPPCs, weighed by the ratio between the feedback resistance  $\mathrm{R_f}$ and $\mathrm{R_s}$. The $\mathrm{R_f}$ value is constrained by the characteristics of the operational amplifier in use, while the $\mathrm{R_s}$ value has to be optimized.

The challenge is to provide a cryogenic readout  that can deal with the capacitance of individual MPPCs, limiting the associated noise  and providing a signal to noise ratio larger than one. 
In Section \ref{sec:detector} the detector used in this experiment is described.
The experimental setup is shown in Section \ref{sec:setup}.
In Section \ref{sec:electronics} we describe the readout.
The results are presented in Section \ref{sec:results}.

\section{The detector}\label{sec:detector}

\begin{figure}[h!]
\centering
\includegraphics[width=0.5\textwidth]{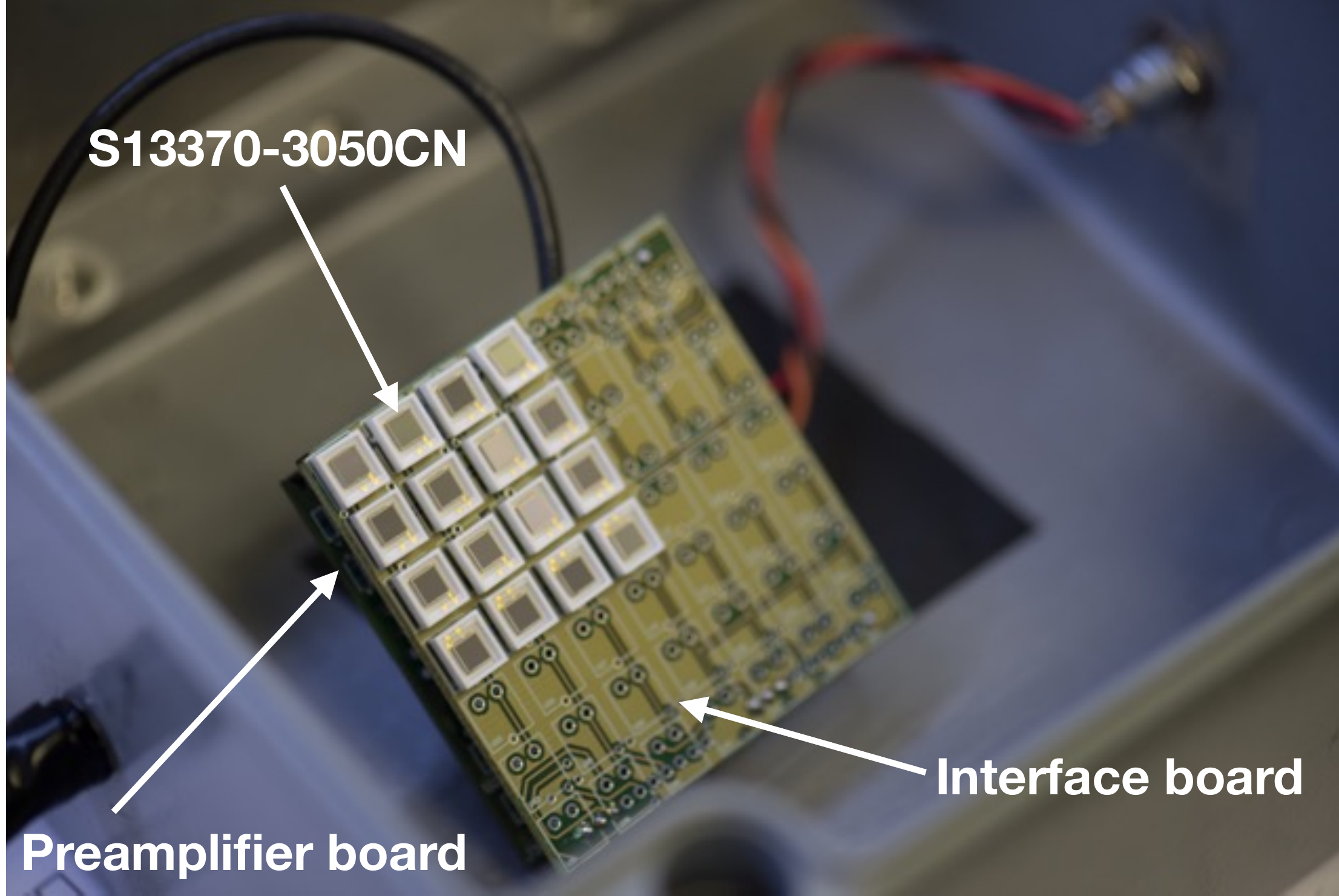}
\caption{\label{fig:detector} The detector with the 16 individual MPPCs used in the experiment. Due to the ceramic package, the maximum number of S13370-3050CN that can fit on the interface board is in fact 49.}
\end{figure}

In Figure \ref{fig:detector}, the prototype of the VUV detector under test is shown. It consists of an array of 16 MPPCs soldered to an interface board that is in turn connected to the preamplifier board. 

The decision to split the readout into interface and preamplifier boards has been taken to have the flexibility of testing different types of MPPCs (individually or grouped in tiles). The electronics is designed to readout up to 64 individual channels; however, due to the ceramic frame of the devices under test, the maximum number of S13370-3050CN that can fit on the interface board is in fact 49. { For the present work we used  16 MPPCs to have a better characterize the electronics under test}.

It is worth mentioning that the AD8011 operational amplifier had been already used by our group, in a preamplifier circuit for the Hamamatsu PMT R11410. The preamplifier was successfully tested at LXe conditions \cite{ref:preamplifierELBA}. 

\section{Experimental Setup}\label{sec:setup}

\begin{figure}[h]
\centering
\includegraphics[width=0.7\textwidth]{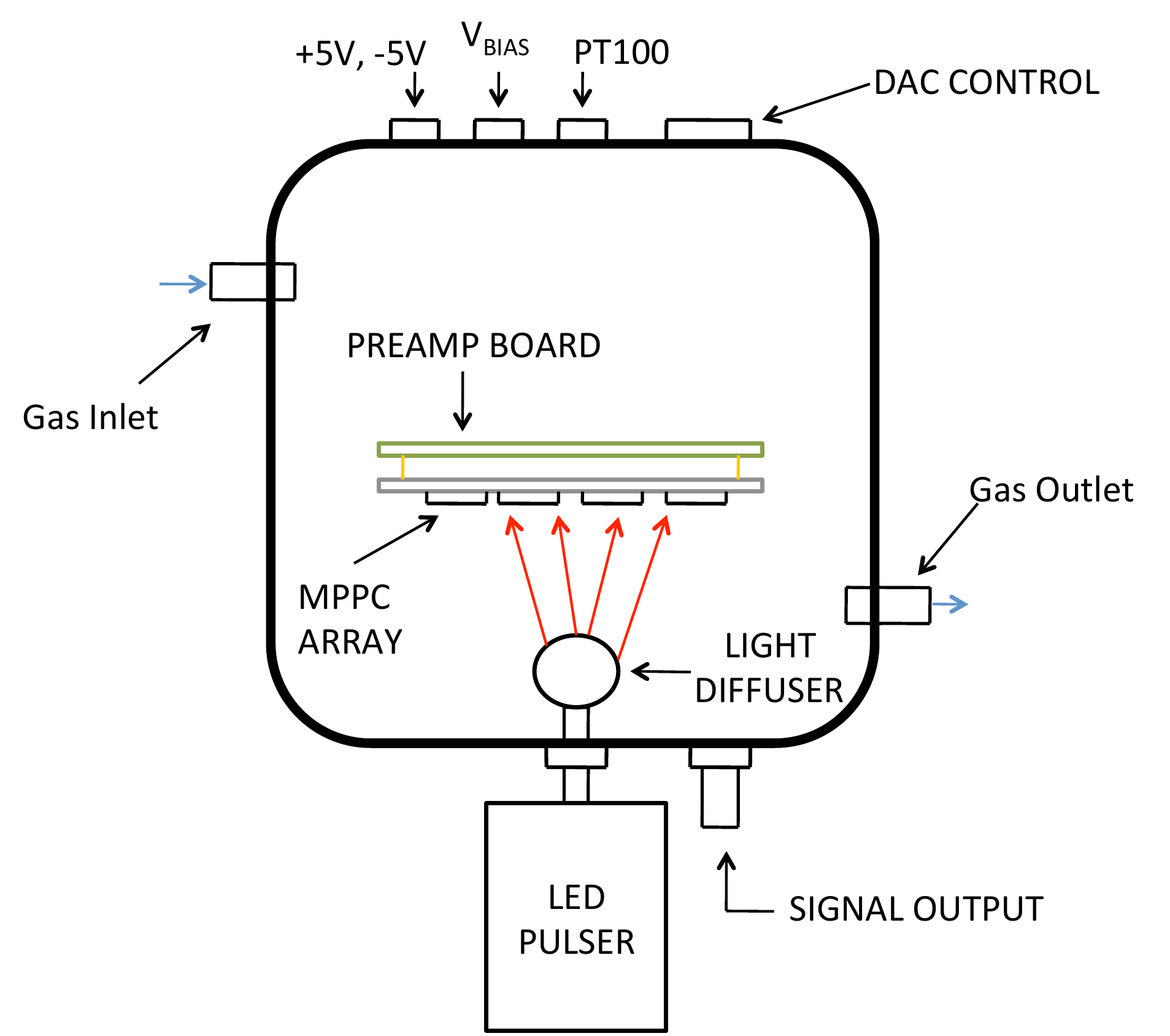}
\caption{\label{fig:aluminumbox} A sketch of the test unit used in the experiment.}
\end{figure}

The characterization of the detector has been performed at cryogenic conditions and more specifically at LXe temperature ($\mathrm{175~K}$) by using a cold finger partially immersed in liquid nitrogen and in direct contact with the setup. Varying the liquid level allows for the control of the temperature. The  MPPC array and its electronics have been placed in an aluminum box (see figure \ref{fig:aluminumbox}) equipped with connectors for  signal readout, for detector and preamplifier biasing, and with an optical diffuser to isotropically deliver light from a pulsed LED towards the sensitive detector surface. To constantly monitor the temperature and to correct the biasing voltage accordingly, a PT100 has been used. Gas nitrogen has been flushed through the box, to avoid discharges arising from water condensation.

To compensate the effect of different breakdown voltages of the 16 MPPCs, (in the range $\mathrm{55.56~V}\div\mathrm{55.78~V}$) the biasing section has been equipped with a Digital to Analog Converter (DAC) to equalize the over-voltages\footnote{ The over-voltage (\ov) is the voltage above the  breakdown}. All the 16 MPPCs have been biased by an Agilent E3645A, while a linear DC Elind 32DP8 power supply has been used to operate the preamplifier. The readout of the temperature through the PT100 has been performed by a Keithley 2100 digital multimeter. A LeCroy HDO6104 high definition oscilloscope has been used for signal readout and data acquisition.   

\begin{figure}[h]
\centering
\includegraphics[width=0.75\textwidth]{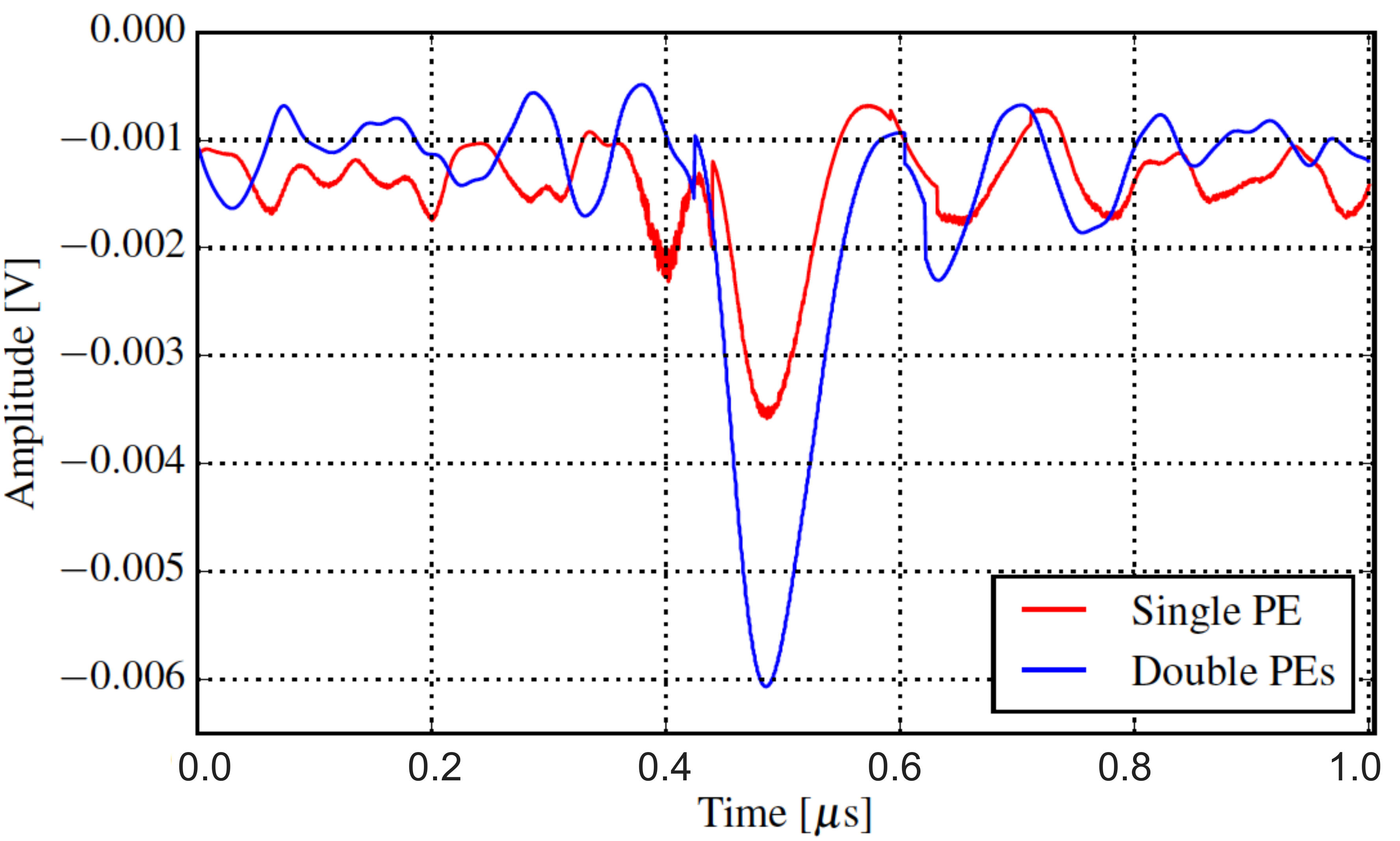}
\caption{\label{fig:waveform} { Example of typical} waveforms corresponding to a single photon and to 2 photons  taken at $\mathrm{175~K}$, \ov$\mathrm{=} \mathrm{2~V}$ ($\mathrm{50~\Omega}$ termination).}
\end{figure}

Each waveform has been collected in $\mathrm{1~\mu s}$ time window and sampled with 2500 points at 12 bits at full bandwidth. All the results shown in this paper are presented without using any noise or off-line filtering. 
 Examples of waveforms corresponding to single and double photon event families are shown in figure \ref{fig:waveform}.

\section{The Electronics}\label{sec:electronics}

\begin{figure}[h!]
\centering
\includegraphics[width=0.8\textwidth]{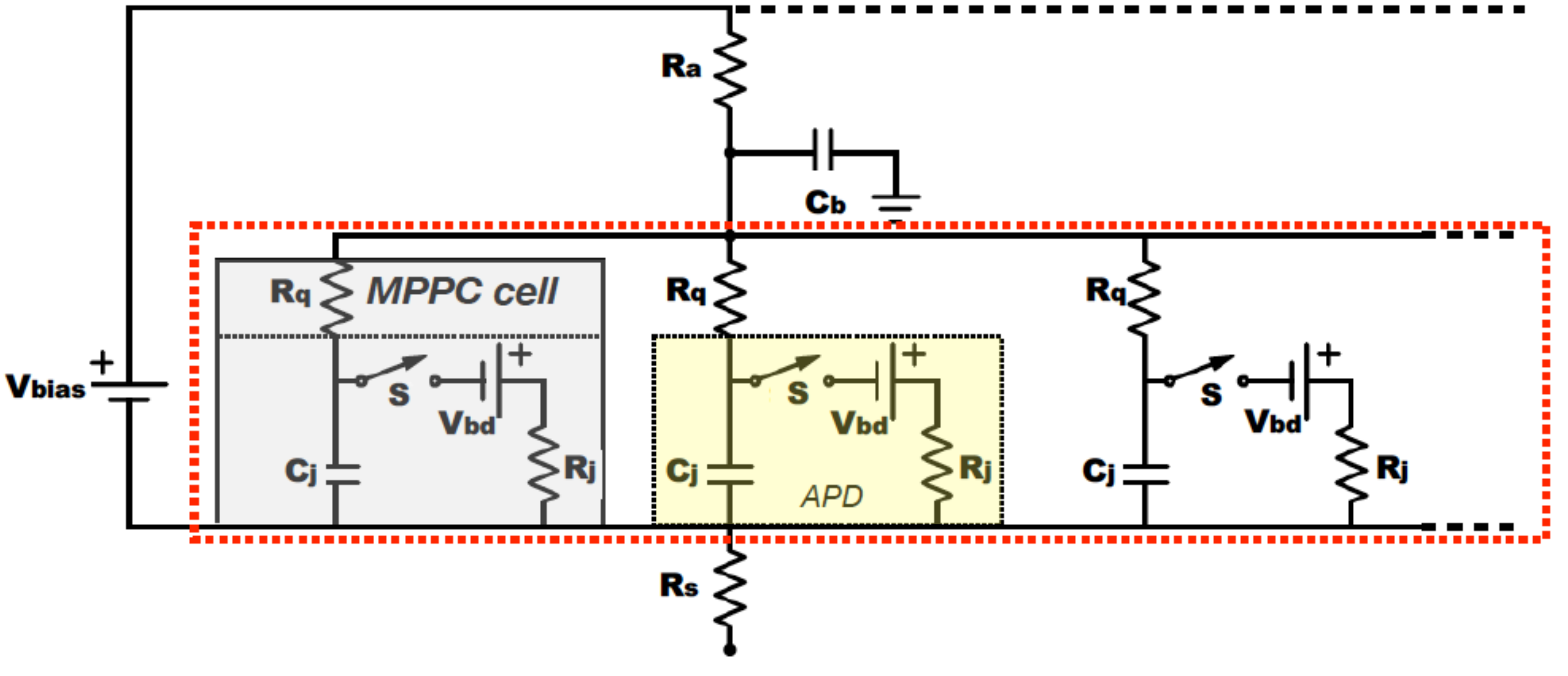}
\caption{\label{fig:MPPCscheme} Equivalent circuit of a MPPC. Each MPPC cell (gray box) is a series connection between a quenching resistor $\mathrm{Rq}$ and an Avalanche Photodiode (APD, yellow box). The red dashed box represents the actual MPPC device consisting of many single MPPC cells connected in parallel.}
\end{figure}

 An MPPC can be modeled as a matrix of many independent channels connected in parallel \cite{ref:hamamatsuMPPC}, each one consisting of a series of one avalanche photodiode (APD) and one quenching resistor ($\mathrm{O(k\Omega)/\DAlambert}$). Figure \ref{fig:MPPCscheme} shows the typical electrical scheme of an MPPC.
 The APD can be represented by a junction capacitance $\mathrm{Cj}$, a voltage source corresponding to the breakdown voltage $\mathrm{Vbd}$, a junction resistance $\mathrm{Rj}$ and a switch S that closes when a photon hits the sensor. The resistor $\mathrm{Rq}$ is then used to quench the signal and restore the APD switch to the open position. Current limiting resistors ($\mathrm{R_a}$), bypass capacitances ($\mathrm{C_b}$) and the decoupling resistors  ($\mathrm{R_S}$) are all wired outside the MPPC.

\begin{figure}[h]
\centering
\includegraphics[width=1.0\textwidth]{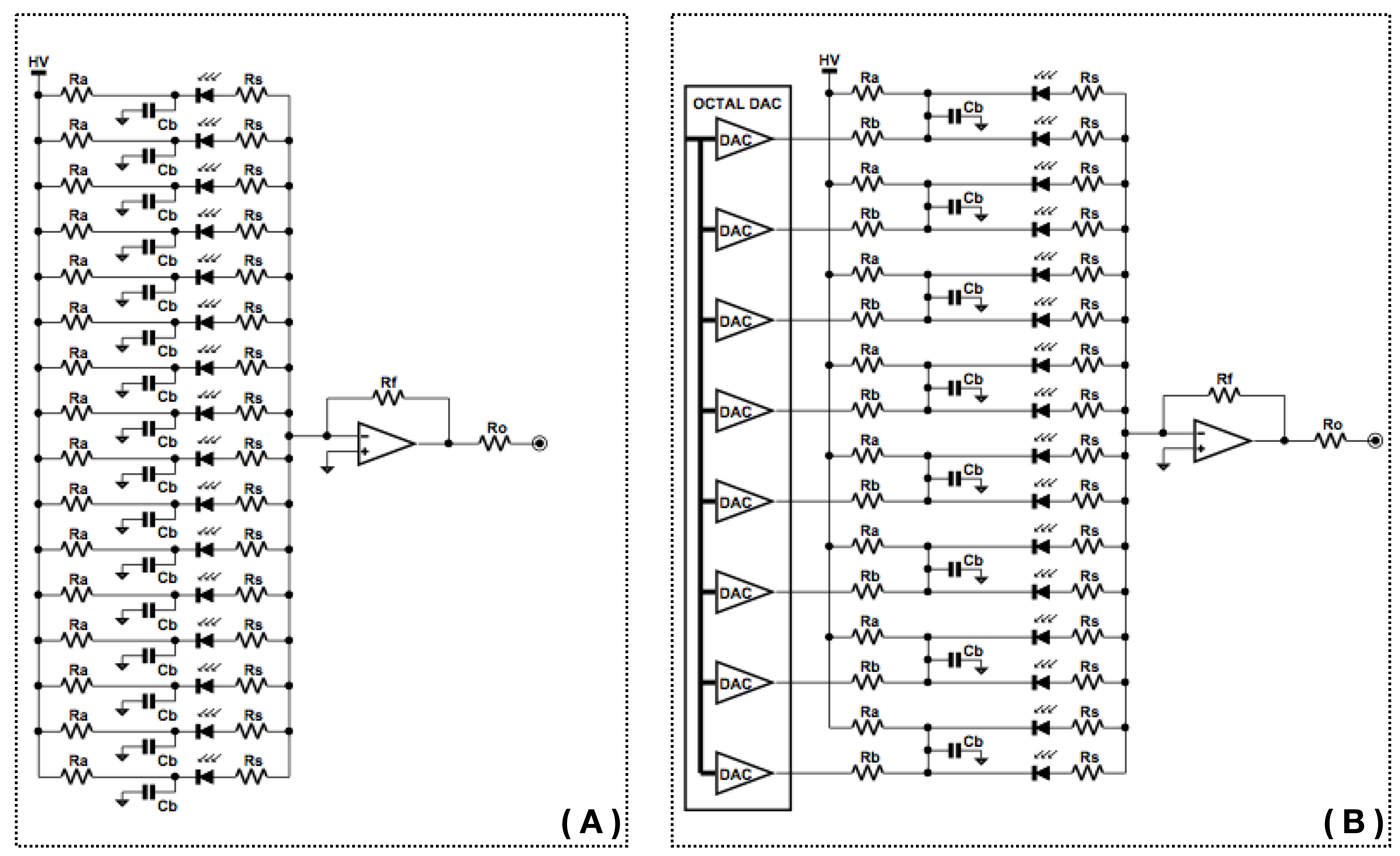}
\caption{\label{fig:Ele_Layout} Layout of the electronics readout. Figure (A) shows the simplest configuration for a 16-channel detector. Figure (B) shows a configuration with the addition of the DAC section used for biasing correction. The analog sum of the 16 signal devices is performed by the AD8011 operational amplifier for both configurations. { $\mathrm{R_a}$ is the current limiting resistor, $\mathrm{C_b}$ is the bypass capacitance, $\mathrm{R_S}$ decouples the MPPC parasitic capacitance.}}
\end{figure}

In Figure \ref{fig:Ele_Layout} two configurations (A and B) for the operation with 16 channels (the electronics can actually host up to 64 devices) are shown.  In A, all  channels are biased with the same voltage, while B  features the additional 8 bit DAC (DAC088S085 by Texas Instruments), suitable for operation at $\mathrm{175~K}$, used to equalize the over-voltage of each channel in steps of $\mathrm{4~mV}$.  Configuration A is suitable when operating arrays where the  breakdown voltage of different cells is highly homogeneous, otherwise configuration B must be used. 

Both circuits have been designed to prevent any contribution of non-fired MPPCs to the analog signal sum of the entire array.
An approach  similar to configuration A was considered for liquid argon based applications, but with a different detector \cite{ref:Sensl,ds-sipm}.

The resistor $\mathrm{R_S}$ is used to decouple the MPPC equivalent parasitic capacitance $\mathrm{C_S}$ of any non-fired photosensor from the operational amplifier. 
The presence of $\mathrm{R_S}$ is what makes this design different from a purely parallel configuration. This technique becomes highly effective at low temperature where the dark counting rate drops dramatically.

\begin{figure}[h!]
\centering
\includegraphics[width=0.7\textwidth]{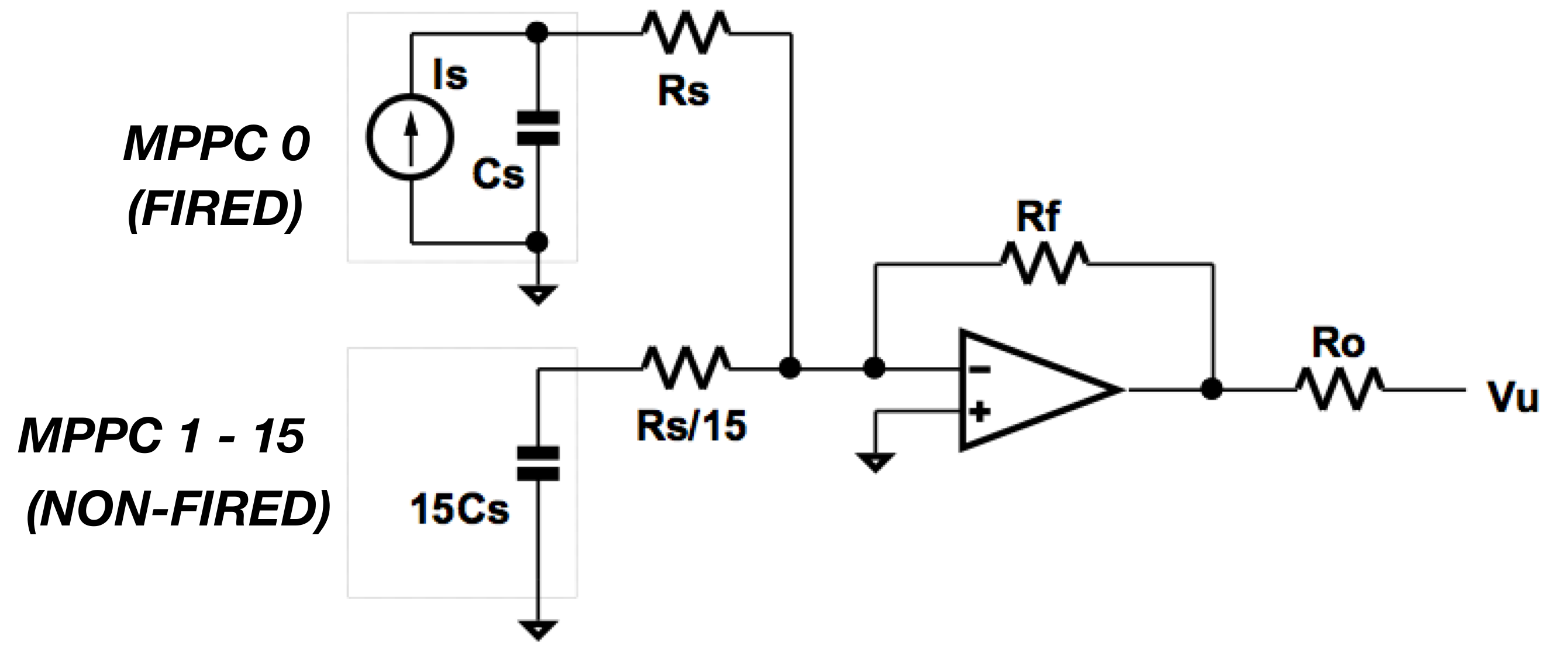}
\caption{\label{fig:FiredMPPC} Scheme of the equivalent circuit of a single fired MPPC.}
\end{figure}

 Figure \ref{fig:FiredMPPC} is an approximate representation of the array when a MPPC (MPPC number 0) detects a photon (is `fired') while the others don't.

The MPPC-0 signal is represented by the current generator $\mathrm{I_S}$ connected in parallel with $\mathrm{C_S}$ towards ground.  $\mathrm{C_S}$ is equivalent to the parallel of the $\mathrm{Cj}$ of each cell composing the MPPC. It is worth mentioning that all the quenching resistors are connected in parallel, resulting in an equivalent resistance that is negligible with respect to $\mathrm{R_S}$. Due to a series connection between $\mathrm{R_S}$ and the equivalent quenching resistance, its effect can be also included in the overall effect of $\mathrm{R_S}$.
The operational amplifier output voltage is, to first order, the product between the value of the feedback resistance  ($\mathrm{R_f}\mathrm{=1~k\Omega}$) and the current $\mathrm{I_S}$, given the negligible effect of $\mathrm{C_S}$ and $\mathrm{R_S}$. 

The average amplitude of a typical single photon event waveform ($\mathrm{175~K,V_{OV}=2~V}$, $\mathrm{50~\Omega~termination}$, see Figure \ref{fig:waveform}) is $\sim$2.5 mV corresponding to $I_S\mathrm{=5~\mu A}$ ($R_f\mathrm{=1~k \Omega}$ and $R_0\mathrm{=50~\Omega}$). 

\subsection{Gain equalization}

The analog sum of signals generated by individual devices operating at different over-voltages will affect the single photon detection capability of the entire array because of the non-uniformity of the gain.  By using the configuration B reported in figure \ref{fig:Ele_Layout}, the  over-voltage of each MPPC group can be fine tuned, resulting in the equalization of the gains, without contributing to the overall noise. Each DAC serves 2 (or more) MPPCs with similar breakdown voltages. Two resistors, $\mathrm{R_A}$ and $\mathrm{R_B}$ are used to distribute, respectively, the biasing voltage and the DAC output.\\
The  biasing voltage is given by the following equation:

\begin{equation}
\mathrm{V_{MPPC}}=\frac{\mathrm{V_{DAC}}\times\mathrm{R_A}}{\mathrm{R_A}+\mathrm{R_B}}+\frac{\mathrm{V_{Bias}}\times\mathrm{R_B}}{\mathrm{R_A}+\mathrm{R_B}}
\label{eq:bias}
\end{equation}

where $\mathrm{V_{MPPC}}$ is the operating voltage of the MPPC, while $\mathrm{V_{DAC}}$ and $\mathrm{V_{Bias}}$ are the output voltages of the DAC and of the power supply. The maximum output voltage of DAC used in the setup is $\mathrm{5~V}$ that allows for a biasing correction in the range of $\pm\mathrm{0.5~V}$. The $\mathrm{R_A}$ and $\mathrm{R_B}$ values are constrained by the equation: 
\begin{equation}
 \frac{\mathrm{R_A}}{\mathrm{R_A}+\mathrm{R_B}}=\frac{1}{5}
\label{eq:correction}
\end{equation}
This configuration preserves the DC coupling of the MPPCs, with a slight increase of the total power budget, as shown below:

\begin{equation}
\mathrm{I_{tot}}=\frac{\mathrm{V_{Bias}-V_{DAC}}}{\mathrm{R_A}+\mathrm{R_B}},~\mathrm{P_{tot}}=\frac{\mathrm{(V_{Bias}-V_{DAC})^2}}{\mathrm{R_A}+\mathrm{R_B}}
\label{eq:powercons}
\end{equation}

Using $\mathrm{R_A}=\mathrm{250~k\Omega}$ and  $\mathrm{R_B}=\mathrm{1~M\Omega}$ in the DAC configuration, the biasing voltage becomes:
\begin{equation}
\mathrm{V_{MPPC}}=\mathrm{V_{DAC}\times 0.2+V_{Bias}\times 0.8}
\label{eq:newbiasing}
\end{equation}

The power consumption can be reduced by increasing the value of the distribution resistors $\mathrm{R_A}~\mathrm{and}~\mathrm{R_B}$, as shown in Table \ref{tab:powerbudget}. 

\begin{table}[h]
\centering
\makegapedcells
\scalebox{0.9}{
\begin{tabular}{|c|c|c|c|c|c|c|c|}
\hline
${\small\mathrm{R_A[M\Omega]}}$ & ${\small\mathrm{R_B[M\Omega]}}$ & ${\small\mathrm{V_{BD}}}\mathrm{+}{\small\mathrm{V_{ov}}}={\small\mathrm{V_{MPPC}~[V]}}$ & ${\small\mathrm{V_{DAC}[V]}}$ & ${\small\mathrm{V_{Bias}[V]}}$ & ${\small\mathrm{P_{tot}[mW]}}$
\\
\hline
0.25         & 1        & 55.5              + 1.0                     = 56.5           & 2.5           & 70             & 3.65\\       
2.5         & 10        & 55.5                + 1.0                     = 56.5           & 2.5           & 70             & 0.36\\       
5         & 20        & 55.5               + 1.0                     = 56.5           & 2.5           & 70             & 0.18\\  
\hline
\end{tabular}}
\caption{Power consumption ($\mathrm{P_{tot}}$) scenarios per channel as a function of the distribution resistors $\mathrm{R_A}\mathrm{and~R_B}$. $\mathrm{V_{BD}}$, $\mathrm{V_{ov}}$, $\mathrm{V_{MPPC}}$, $\mathrm{V_{DAC}}$ and $\mathrm{V_{Bias}}$ are the MPPC breakdown voltage, the over-voltage, the voltage across the MPPC, the DAC output and the biasing voltage respectively.}\label{tab:powerbudget}

\end{table}

The number of DACs used in the proposed configuration can be reduced  if the MPPC  breakdown voltage spread is small enough.  The more MPPCs connected to a single DAC, the lower  the overall power absorbed. In the present configuration, each DAC serves 2 MPPCs only, but considering the narrow span of breakdown voltages of the sample, more devices could be connected to a single DAC channel.

\subsection{Noise Model}
\label{sec:NoiseModel}

\begin{figure}[h!]
\centering
\includegraphics[width=0.7\textwidth]{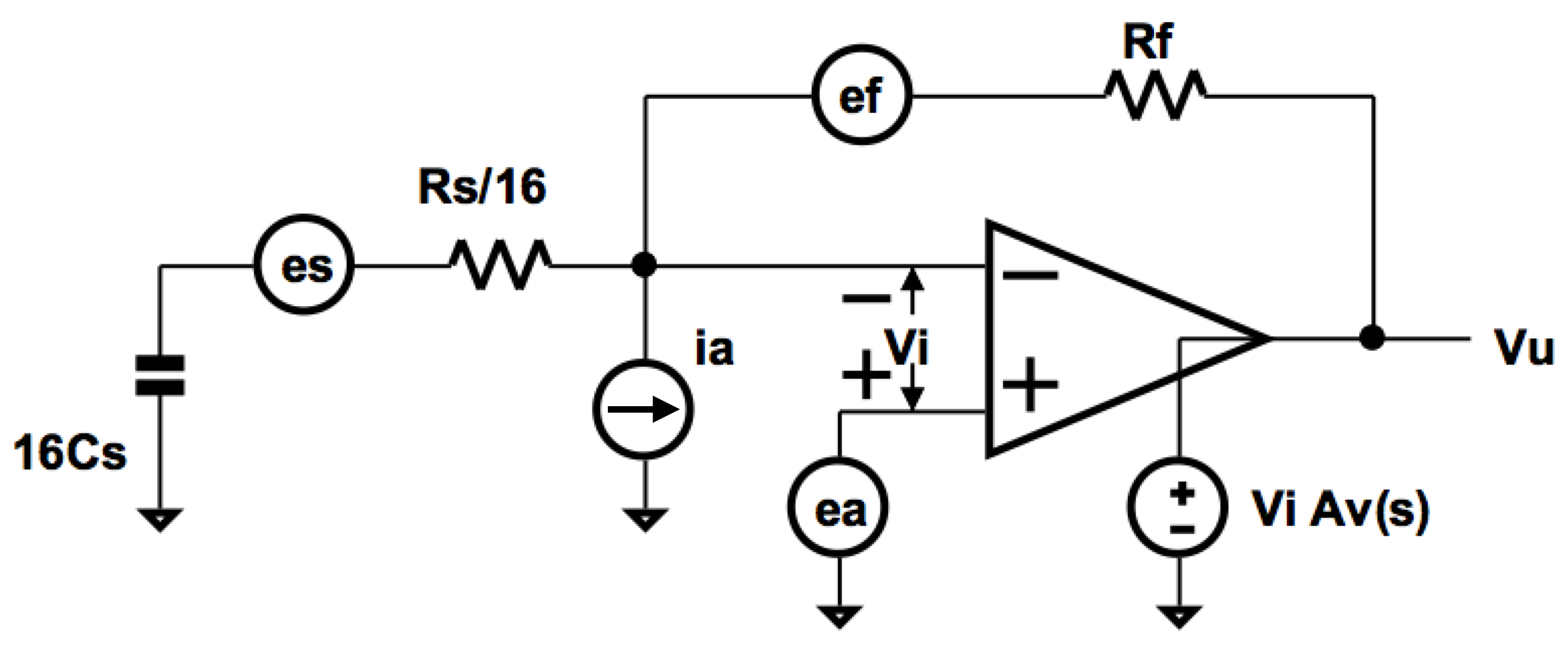}
\caption{\label{fig:NoiseSources} Schematics of the circuit for 16 MPPC connected devices showing the noise sources. Generators $\mathrm{e_s}$ and $\mathrm{e_f}$ are the noise voltage contributions of resistors $\mathrm{R_s}$ and $\mathrm{R_f}$. $\mathrm{C_s}$ is the parasitic capacitance of a MPPC ($\mathrm{C_s}=\mathrm{320~pF}$ at room temperature) and $\mathrm{R_s}$ is the connection resistor ($\mathrm{R_s}=\mathrm{50~\Omega}$ at room temperature). Generators $\mathrm{e_a}$ and $\mathrm{i_a}$ are the input noise voltage and current of the AD8011. $\mathrm{V_i}$ and  $\mathrm{V_u}$ are the input and output operational amplifier voltages, while $\mathrm{A_v(s)}$ is the amplifier open loop gain value.}
\end{figure}

The non-correlated noise sources considered in the design of the circuit are shown in Figure \ref{fig:NoiseSources}. The AD8011 is a current feedback operational amplifier, however the proposed circuit has been studied and operated as a transimpedance amplifier (TIA). This assumption is considered valid, as reported in  \cite{ref:TexasInstruments}.  $\mathrm{e_s}$ and $\mathrm{e_f}$ represent the noise voltage contributions of resistors $\mathrm{R_s}$ and $\mathrm{R_f}$. $\mathrm{C_s}$ is the parasitic capacitance of a MPPC. $\mathrm{e_a}$ and $\mathrm{i_a}$ are the input noise voltage and current of the AD8011. The input and output operational amplifier voltages are $\mathrm{V_i}$ and $\mathrm{V_u}$, respectively, while $\mathrm{A_{v(s)}}$ is the amplifier open loop gain value.

The impedance of all the input branches, expressed in terms of the complex variable $\mathrm{s}$ in the Laplace transform space, is given by the following equation:

\begin{equation}
\mathrm{Z_s(s)}=\frac{\mathrm{1+s \times R_s \times C_s}}{\mathrm{16 \times s \times C_s}}
\label{eq:branchimpedance}
\end{equation}

To account for the overall noise contribution, the transfer functions of each uncorrelated noise source have been evaluated, and added  quadratically. To estimate the total noise effect, we run a simulation  assuming infinite input impedance and  open loop gain $\mathrm{A_v(s)}$ with finite value and dominant pole. The system has been modeled as a voltage-feedback amplifier for the analysis of the total noise.\\
The transfer function of each noise source in the Laplace space can be written as: 
\begin{equation}
\frac{\mathrm{V_u(s)}}{\mathrm{k}}=\mathrm{B(k,s)} \times \mathrm{f(s)}
\label{eq:transfGeneric}
\end{equation}
where $\mathrm{k}$ is the noise source ($\mathrm{i_a,~e_s,~e_f,~e_a}$), $\mathrm{B}$ is a factor specific for each source and $\mathrm{f(s)}$ is a common component to all the transfer functions.

 Each transfer functions can be expressed in the frequency domain by replacing the Laplace variable with the complex expression $\mathrm{j2\pi f}$, where f is the frequency.

$\mathrm{B(k,f)}$ for each noise source is reported in Table \ref{tab:noise}. The contribution of the input voltage noise $\mathrm{e_a}$ is the most significant.

\begin{table}[h]
\centering
\makegapedcells
\scalebox{0.75}{
\begin{tabular}{|c|c|c|c|c|c|}
\hline
$\mathbf{Source~of~Noise}$ & $\mathrm{i_a}$ & $\mathrm{e_a}$ & $\mathrm{e_f}$ & $\mathrm{e_s}$ 
\\
\hline

$\mathbf{B(k,f)}$         &$i_a\times\mathrm{R_f}$        & $e_a\times (1+\frac{\mathrm{R_f}}{\mathrm{Z_s(f)}})$               & $4\mathrm{KT} \times R_f$                  & $4 \mathrm{KT} \frac{R_f}{Z_s(f)} $  
\\
\hline
$\mathbf{\frac{Spectral~density~noise}{C(f)}}~~\mathbf{[\frac{V}{\sqrt{Hz}}]}$         &$\mathrm{5.0\times 10^{-9}}$        & $\mathrm{\le 6.4\times 10^{-7}}$               & $\mathrm{9.6\times 10^{-18}}$ & $\mathrm{\le~3\times 10^{-18}}$  \\ 
\hline
\end{tabular}}
\caption{Noise spectral density contributions to the transfer function. The main noise component is the amplifier input voltage noise $\mathrm{e_a}$.  $\mathrm{K}$ and $\mathrm{T}$ are the Boltzmann constant and the absolute temperature respectively. The contributions $\mathrm{e_s}$ and $\mathrm{e_a}$ have been maximized considering the frequency dependence of $\mathrm{Z_s}$.} 
\label{tab:noise}
\end{table}

 The simulation has been performed by transforming  the transfer functions from the Laplace s-domain to the z-transform domain resulting in a recursive equation in the time domain. The solution of the equation, once set the boundary conditions, is a waveform that can be compared to the measured one. Due to the intrinsically random process, simulated and measured waveforms should be only compared qualitatively.

\begin{figure}[h!]
\centering
\includegraphics[width=0.8\textwidth]{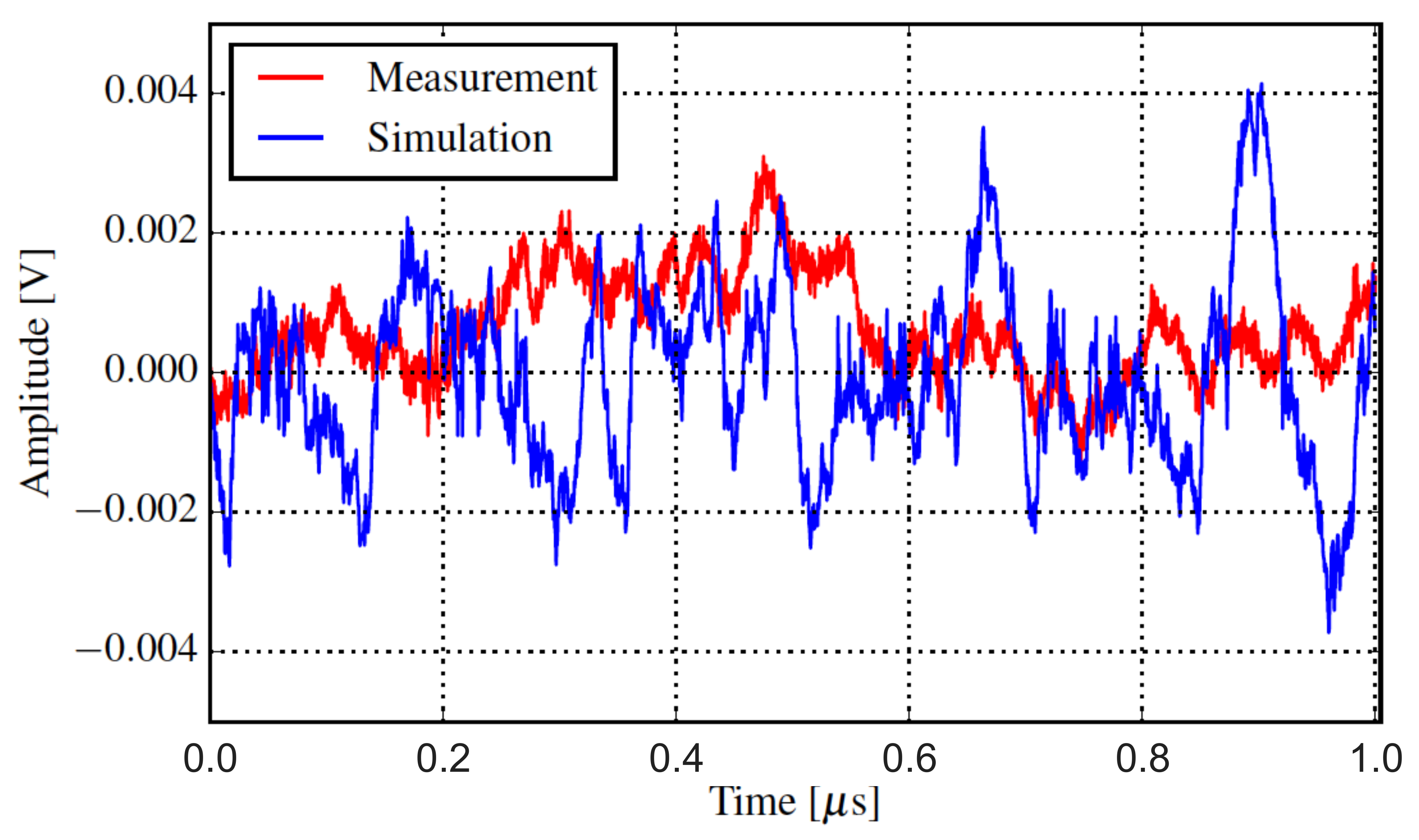}
\caption{\label{fig:noisesimwave} Example of simulated and acquired noise signals.}
\end{figure}

{ Time domain} simulated and acquired noise waveforms (taken at room temperature) are shown in Figure \ref{fig:noisesimwave}. { The RMS evaluated for real waveforms, over a sample of 10,000 events taken at $\mathrm{175~K}$, is of the order of $\mathrm{900 ~\mu V}$. 
The RMS of simulated waveforms is  $\mathrm{660~\mu V}$.}

\section{Results}
\label{sec:results}
\begin{figure}[h]
\centering
\includegraphics[width=0.9\textwidth]{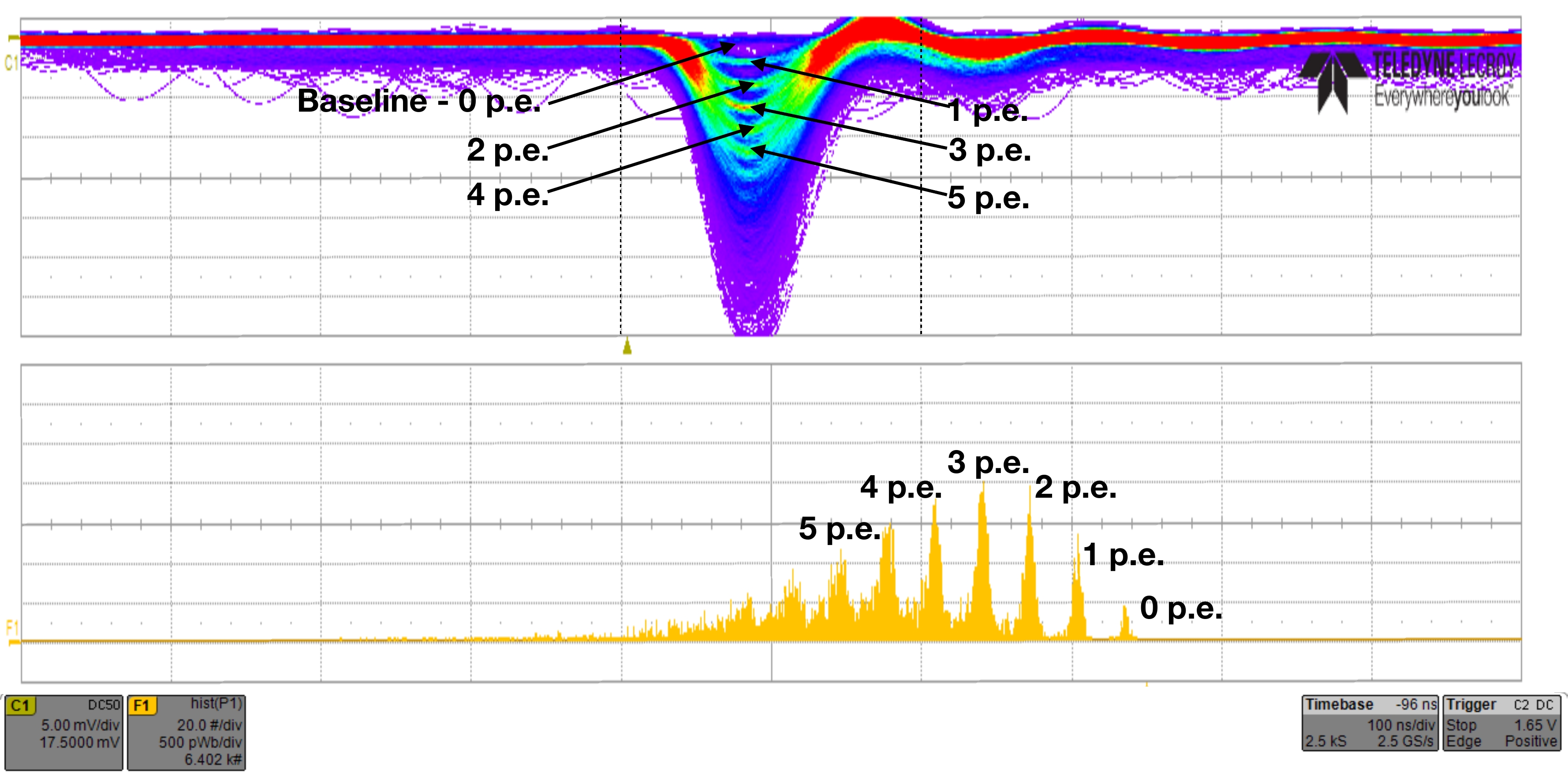}
\caption{\label{fig:SinglePEseparation} (Top) Waveforms taken in persistence mode at $\mathrm{175~K}$, $\mathrm{V_{OV}=3~V}$, by illuminating the detector with a LED pulser (used as trigger too). The spacing between signal families is the consequence of the preserved single photon counting capability after summing up the 16 individual MPPCs. The baseline and the first 5 photoelectrons (p.e.) families are tagged,  as well as the time window used for the waveform integration (Bottom). The bottom spectrum should be read from right to left.}
\end{figure}

\begin{figure}[h]
\centering
\includegraphics[width=0.7\textwidth]{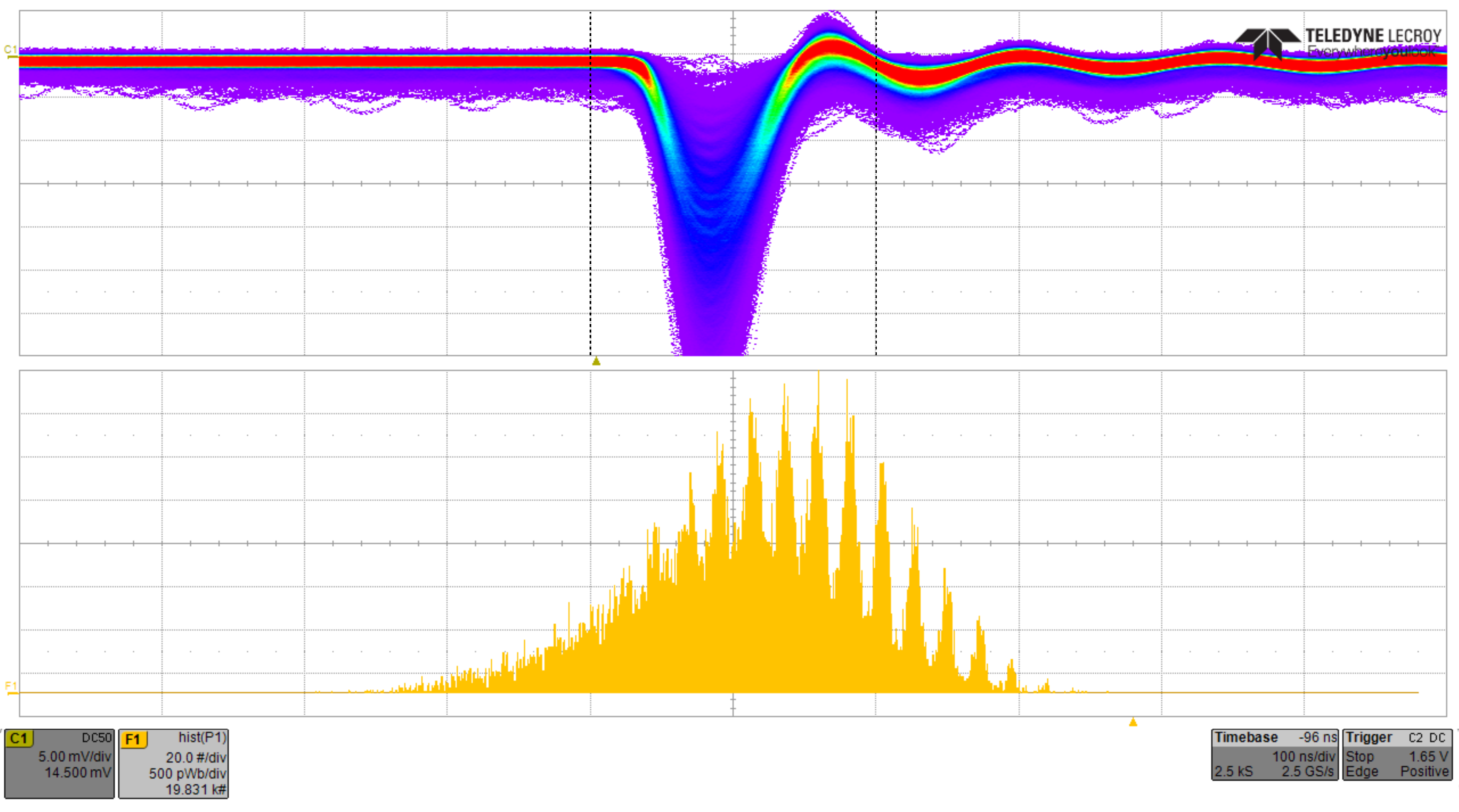}
\caption{\label{fig:SinglePEcapability} Waveforms (top) in persistence mode and p.e. spectrum (bottom) taken at $\mathrm{175~K}$, $\mathrm{2~V}$ of over-voltage, by illuminating the detector with a LED pulser (used as trigger too). Families of signals, corresponding to single photons are clearly visible. The integration of signals produces the typical peaked photoelectron spectrum. The bottom spectrum should be read from right to left.}
\end{figure}

To test the single photon detection capability of our system, we have operated the array (with 16 MPPCs) in configuration A (see figure \ref{fig:Ele_Layout}), $\mathrm{T=175~K}$, $\mathrm{V_{OV}=3~V}$ and illuminated by a pulsed UV LED.
The measurements have been performed at low intensity to maximize the probability of having only one photon per event (see figure  \ref{fig:SinglePEseparation}) and at a higher intensity for the detection of several photons  per event (see figure \ref{fig:SinglePEcapability}).
The separations of signal families corresponding to different number of photons detected in a single event is shown in figure \ref{fig:SinglePEseparation}, where the waveforms have been acquired in persistence mode. 
The first family, below the baseline band, corresponds to events triggered by single photons, the second family to events triggered by two photons, etc.  
The corresponding integrated values in the $\mathrm{200~ns}$ window centered around the signal peak are shown in figure \ref{fig:SinglePEseparation} (bottom).

\begin{figure}[h]
\centering
\includegraphics[width=1.\textwidth]{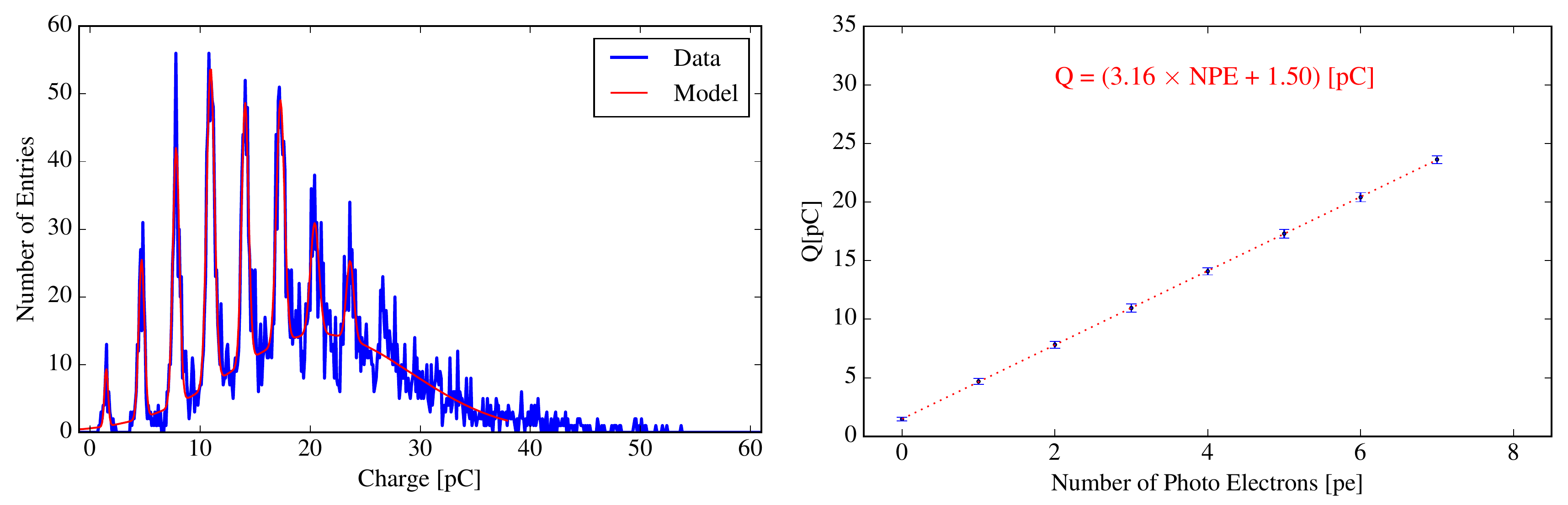}
\caption{\label{fig:SinglePEFIT} Photoelectron charge distribution with the array operated at $\mathrm{175~K}$, $\mathrm{V_{OV}=3~V}$. The data have been fitted by using a set of multiple gaussians. The measured charge of the 0 p.e. peak (baseline) is of the order of $\mathrm{1.5~pC}$, while the measured charge of the single photoelectron peak is about $\mathrm{3.2~pC}$.}
\end{figure}

The photoelectron spectrum has been therefore fitted with a set of multiple gaussians (see figure \ref{fig:SinglePEFIT}): the measured charge of the pedestal is $\mathrm{(1.47 \pm 0.16) ~pC}$, while the single photoelectron peak is found at about $\mathrm{(3.21 \pm 0.26)~pC}$, corresponding to an overall gain of $\sim \mathrm{2.0 \times 10^7}$ ($\mathrm{175~K}$, \ov $=\mathrm{3~V}$). 

{ Since the sigma of the charge pedestal showed in figure \ref{fig:SinglePEFIT} is smaller than the average charge separation between two consecutive peaks ($\mathrm{0.16~ pC}$ versus $\mathrm{3.16~ pC}$),  a distinctive photoelectron peak charge distribution can be observed.  Assuming no statistical effects, the main contributions to the sigma of each photoelectron peak are in fact due to noise ($\sigma_{ELE}$), dark counts ($\sigma_{DC}$), afterpulses ($\sigma_{AP}$), crosstalk ($\sigma_{CT}$) and gain fluctuation ($\sigma_{GF}$): the corresponding integrated charge of any signal emerging from the tail of a spurious event (AP, CT or DC) is slightly overestimated if compared to a signal rising from the baseline.

The equation:
\begin{equation}
\sigma^2_{p.e.}=\sigma^2_{ELE}+\sigma^2_{DC}+\sigma^2_{AP}+\sigma^2_{CT}+\sigma^2_{GF}
\label{eq:sigmas}
\end{equation}
Since  $\mathrm{\sigma_{p.e.}}$ of the first photoelectron peak is $\mathrm{0.26~ pC}$ and  $\sigma_{ELE}$ is $\mathrm{0.16~ pC}$, the contributions due to the detector (AP+DC+CT+GF) is dominant ($\mathrm{0.20~ pC}$). By improving the gain uniformity (using configuration B in figure \ref{fig:Ele_Layout} and the MPPC characteristics), we could conclude that the maximum number of MPPCs used in the array and readout as a single channel can be increased.}

Figure \ref{fig:SinglePEcapability} (top) shows waveforms acquired in persistence mode when operating the LED at high intensity, $\mathrm{T=175~K}$ and \ov $=\mathrm{2~V}$.

In comparison with figure \ref{fig:SinglePEseparation}, a  higher number of families can be identified and their separation is still preserved. The charge spectrum of the acquired signals is shown in figure \ref{fig:SinglePEcapability} (bottom). 

\begin{figure}[h]
\centering
\includegraphics[width=1.\textwidth]{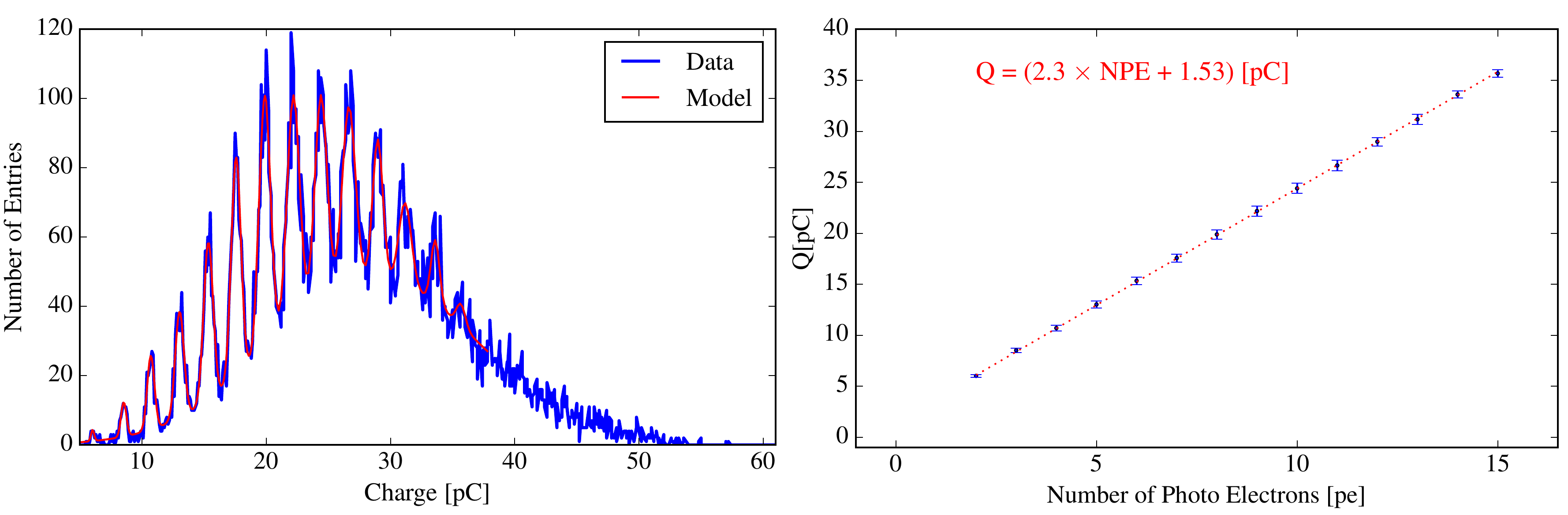}
\caption{\label{fig:GaussianFit} Left: Fit of the charge spectrum, 14 peaks in total have been identified. Right: linearity plot. Its slope estimates the charge of the single photon, while the intercept gives the overall noise average charge.}
\end{figure}

To quantify the linearity of the device, a fit with a set of 14 Gaussian functions has been applied to the acquired data (Figure \ref{fig:GaussianFit}, left). 

The charge value of each peak is reported as a function of the number of detected photons  (see Figure \ref{fig:GaussianFit}, right). The gain of the photodetector  while the the average charge separation between to consecutive photopeaks is $\sim \mathrm{2.3~pC}$, the corresponding gain is $\sim \mathrm{1.4 \times 10^7}$.

\subsection{$^\mathrm{241}$Am spectrum using a LYSO crystal.}

\begin{figure}[h]
\centering
\includegraphics[width=.75\textwidth]{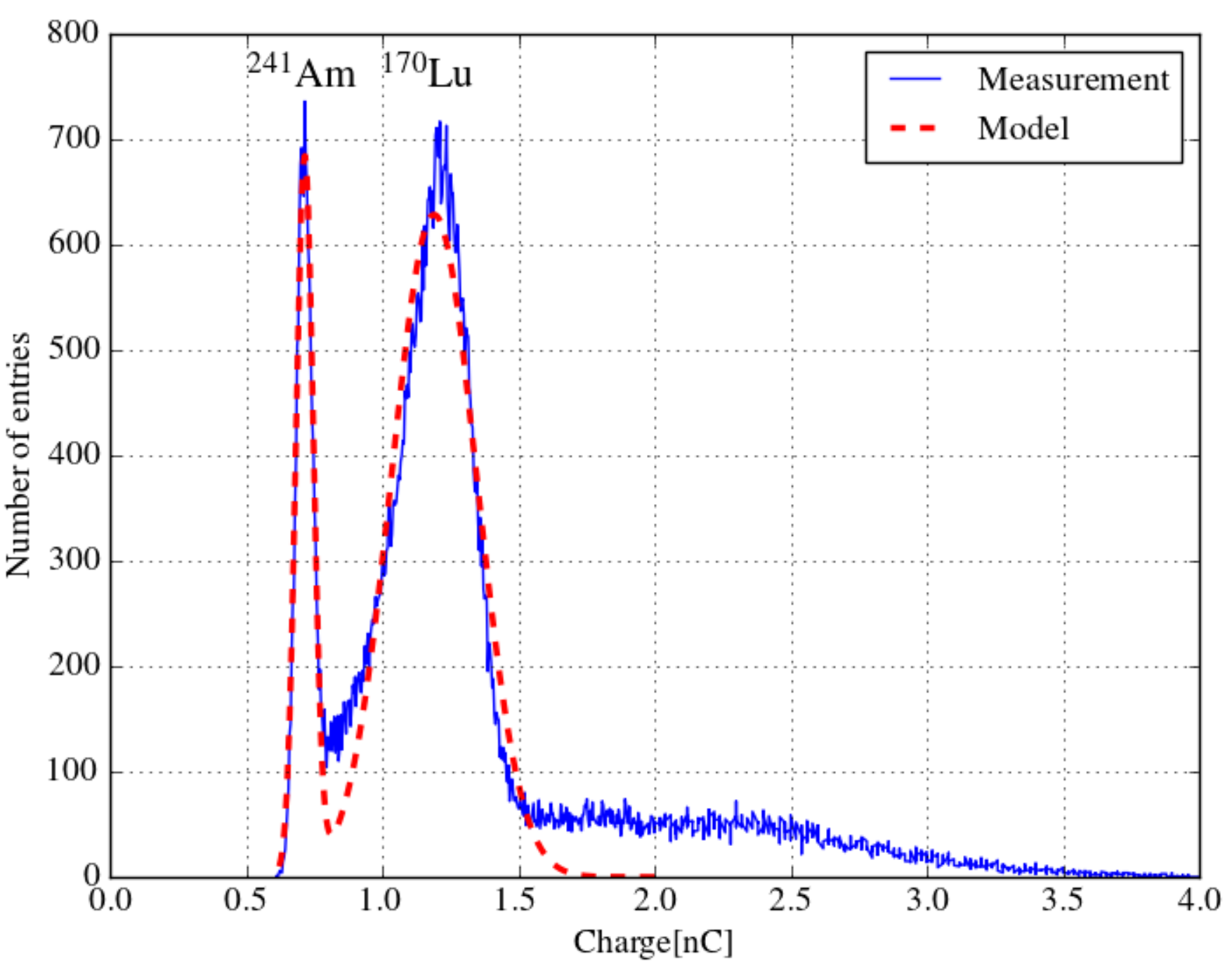}
\caption{\label{fig:241Am} Charge spectrum of $^\mathrm{241}$Am measured by using a LYSO crystal coupled to the MPPC array. The fit of the 59.6 keV line gives the charge value $\mathrm{720~pC \pm 30~pC }$, corresponding to $\mathrm{11~\%}$ resolution.}
\end{figure}

To evaluate its spectroscopic capabilities, the array was coupled to a $\mathrm{15 \times 15 \times50~mm^3}$ Lutetium-Yttrium Oxyorthosilicate (LYSO) crystal irradiated by a $^\mathrm{241}$Am source ($\mathrm{175~K}$, $ V_{OV}=\mathrm{2~V}$). The charge spectrum is shown in figure \ref{fig:241Am}, where the 59.6~keV $^\mathrm{241}$Am gamma peak and the intrinsic LYSO background spectrum due to $\mathrm{^{176}Lu}$ are clearly visible.

{The measured  charge in correspondence of the 59.6 keV line  is $ \mathrm{720~pC\pm 30~pC }$ (corresponding to  about $ \mathrm{310~photons}$). The LYSO light yield is about 30 photons/keV \cite{ref:LYSO}, that is about $ \mathrm{1800~photons}$ for the 59.6 keV gamma emission from $^\mathrm{241}$Am.} Taking into account the photo-electron conversion efficiency, the geometrical coverage of the crystal and non optimal crystal coupling, an overall photon detection efficiency of the order of $\mathrm{17 \%}$ is plausible.
{The estimated energy resolution at the 59.6 keV line is $\sim 11\%$}.

\section{Conclusions}

The aim of this work is to provide LXe detectors, searching  for WIMP-nucleus interactions, the possibility of using MPPCs for the direct detection of scintillation VUV light.

The array, made of 16 individual Hamamatsu VUV4 MPPCs (S13370-3050CN), operates as a single detector by means of a board  based on an operational amplifier suitable for cryogenic environments (AD8011). The total power consumption is about  $\mathrm{10 ~mW}$ per array. 

The electronics foresees a configuration based on the use of Digital to Analog Converter (DAC). This configuration is  more demanding in terms of power supply ($\mathrm{3 ~mW}$ per DAC channel), but it increases the single photon counting capability by compensating the gain differences between MPPCs operating at the same biasing voltage.   

The  noise analysis demonstrated that the only relevant source of noise for the proposed electronics is given by the voltage noise of the input of the operational amplifier.  

The measurements show an excellent photo detection capability. Overall, the performance of this device seems to be promising enough to warrant further studies for its use in liquid xenon based detectors.

\end{document}